\documentclass[epj
]{svjour}
\usepackage{amsfonts,bm,amsmath,amssymb}
\def\ba{{\bm a}}   
 \def\br{{\bm r}} 
\def\bC{{\bm C}} \def\bD{{\bm D}} 
  \def\bL{{\bm L}}
\def\bP{{\bm P}} \def\bS{{\bm S}} \def\bT{{\bm T}} \def\bZ{{\bm Z}}

\def\cA{{\cal A}} \def\cC{{\cal C}}  \def\cR{{\cal R}}
  \def\cZ{{\cal Z}}

\def\rin{\mathrm{in}}\def\rout{\mathrm{out}}
\def\ri{\mathrm{i}}

\newcommand{\fracnul}[2]{\genfrac{}{}{0pt}{}{#1}{#2}}
\def\emline#1#2#3#4#5#6{%
       \put(#1,#2){\special{em:moveto}}%
       \put(#4,#5){\special{em:lineto}}}
\def\newpic#1{}

\begin{document}
 \title{Symmetry Based Properties of the
Transition Metal Dichalcogenide Nanotubes}

 \author{I. Milo\v sevi\'c\thanks{\emph{E-mail:} ivag@afrodita.rcub.bg.ac.yu},
 T. Vukovi\'c, M. Damnjanovi\'c and B. Nikoli\'c}
 \authorrunning{I. Milo\v sevi\'c {\sl et al.}}
 \titlerunning{Transition Metal Dichalcogenide Nanotubes}

\offprints{I. Milo\v sevi\'c}          
\institute{Faculty of Physics, University of Belgrade, P O Box 368,
YU-11001 Belgrade, http://www.ff.bg.ac.yu/qmf/qsg\_e.htm}
\date{13 April 2000}
%
\abstract{The full geometrical symmetry groups (the line groups) of
the monolayered, 2Hb and 3R polytypes of the inorganic MoS$_2$ and
WS$_2$ micro- and nanotubes of arbitrary chirality are found. This is
used to find the coordinates of the representative atoms sufficient
to determine completely geometrical structure of tubes. Then some
physical properties which can be deduced from the symmetry are
discussed: electron band degeneracies, selection rules, general forms
of the second rank tensors and potentials, phonon spectra.
} 
\maketitle
%

\section{Introduction}
Layered structures of the transition metal dichalcogenides MS$_2$
(M=Mo,W) are unstable against folding \cite{WILSON} and this can lead
to a formation of nanotubes \cite{TENNE} or microtubes \cite{R68}.
The obtained tubes have high symmetry, which is relevant both for
profound insight into their physical properties and for
simplification of the calculations. The aim of this paper is to find
the symmetry of ideal single-layered rhombohedral 3R and hexagonal
2Hb polytypes of MS$_2$ \cite{WILSON} folded into tubes, and to
discuss their general symmetry properties.

In Sect.~\ref{SYMMETRY}, after a brief reminder on the line groups,
the symmetry groups for all the chiralities of the monolayered (i.e.
consisting of only one sulfur--transition metal--sulfur molecular
layer), 2Hb and 3R single-layered polytypes of the MS$_2$ nanotubes
are determined. A convenient choice of the lattice vectors enables to
perform this task in analogy with single-wall carbon nanotubes
\cite{YITR}. It turns out that the obtained symmetry groups are
enough to generate the tubes from only 3, 6 and 9 atoms,
respectively; the coordinates of these atoms are given, thus
completing determination of the tubes conformations. In
Sect.~\ref{PROPERTIES} some of many symmetry based physical
properties of the nanotubes are discussed: conserved quantum numbers
and band degeneracies, normal modes, second-rank tensors and
potentials.

\section{Symmetry of the MS$_2$ nanotubes}\label{SYMMETRY}
When  molecular monolayer exhibiting trigonal symmetry is folded up,
the obtained tube is a quasi-1D structure translationally periodic in
the direction of tube axis and its Euclidean symmetry is described by
the line group \cite{LG}, as well as the reported 3R and 2Hb
polytypes tubes, built of two and three S-M-S layers. Therefore,
after a brief review on the line groups, the geometry of the
structures (although still not synthesized) formed by rolling up a
monolayer is considered first.

\subsection{Line groups}
Any quasi-1D system being translationally periodical (with period
$a$) along one axis ($z$ axis, by convention) is a regular
arrangement of elementary structural units conventionally called
monomers. This arrangement is achieved by acting on one of the
monomers by the operations forming generalized translational group
$\bZ$, which is either screw axis $\bT_q^r(a)$ (here belongs pure
translational group $\bT^0_q(a)$) or glide plane $\bT_c(a)$ group.
Also, the monomer itself may have some point group symmetry
\cite{ALTMANN-2}. Only those point operations leaving $z$-axis
invariant are compatible with the 1D translational periodicity,
meaning that the maximal axial subgroup $\bP$ of the monomer point
group should be combined with $\bZ$ to get full symmetry line group.
Namely, the candidates for $\bP$ are groups $\bC_n$, $\bS_{2n}$,
$\bC_{nh}$, $\bC_{nv}$, $\bD_n$, $\bD_{nd}$, $\bD_{nh}$, where
$n=1,2,\dots$ is the principal rotational axis order. Altogether
there are 13 infinite families of the line groups $\bL=\bZ\bP$
differing in type of either $\bZ$ or $\bP$, and with $n$, $q$ and $r$
enumerating groups within each family. This means that each element
of a line group is of the form $g=z^tC^s_nP$ ($t=0,\pm1,\dots$,
$s=0,1,\dots,n-1$), where the screw axis $\bT^r_q(a)$ is generated by
$z=(C_q^r|\frac{na}{q})$ (Koster-Seitz notation), while in the glide
plane case $z=(\sigma_v|\frac{a}{2})$; $C_n$ is principle axis
rotation of the point group and $P$ is some of the remaining point
group operations $U$ (the rotation for $\pi$ around horizontal axis),
$\sigma_v$ and $\sigma_h$ (vertical and horizontal mirror planes).
Monomer is only a part of an elementary cell: the translational
period of $\bT^r_q(a)\bP$ and $\bT_c(a)\bP$ contains $q/n$ and 2
monomers, respectively.

Note that the value of $r$ is not unique: together with $r$ also
$r+m\frac{q}{n}$, $m=\pm 1, \pm 2,\dots$ lead to the same line group.
The ambiguity is resolved by convention choosing $r$ to be coprime
with $q$ (alternatively,  the minimal $r$ may be used, when it is
coprime to $q/n$).

\subsection{Monolayered tubes}\label{Smono}
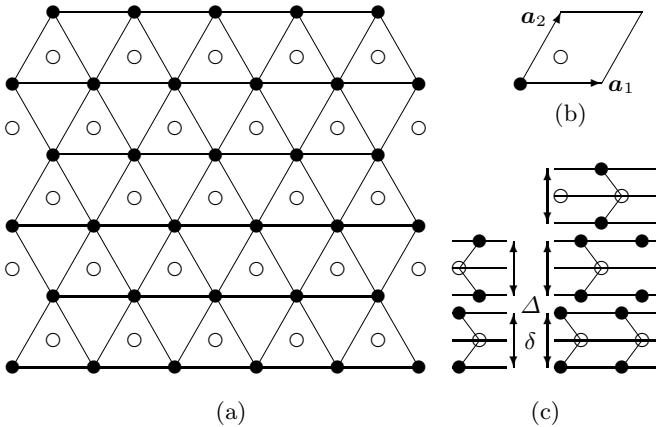
\begin{figure}[hbt]
\unitlength=0.90mm \linethickness{0.4pt}
\begin{picture}(98.07,61.63)
\put(3.07,8.13){\circle*{2.00}} \put(15.07,8.13){\circle*{2.00}}
\put(27.07,8.13){\circle*{2.00}} \put(39.07,8.13){\circle*{2.00}}
\put(51.07,8.13){\circle*{2.00}} \put(63.07,8.13){\circle*{2.00}}
\
\put(9.07,18.63){\circle*{2.00}} \put(21.07,18.63){\circle*{2.00}}
\put(33.07,18.63){\circle*{2.00}} \put(45.07,18.63){\circle*{2.00}}
\put(57.07,18.63){\circle*{2.00}}
\
\put(3.07,29.13){\circle*{2.00}} \put(15.07,29.13){\circle*{2.00}}
\put(27.07,29.13){\circle*{2.00}} \put(39.07,29.13){\circle*{2.00}}
\put(51.07,29.13){\circle*{2.00}} \put(63.07,29.13){\circle*{2.00}}
\
\put(9.07,39.63){\circle*{2.00}} \put(21.07,39.63){\circle*{2.00}}
\put(33.07,39.63){\circle*{2.00}} \put(45.07,39.63){\circle*{2.00}}
\put(57.07,39.63){\circle*{2.00}}
\
\put(3.07,50.13){\circle*{2.00}} \put(15.07,50.13){\circle*{2.00}}
\put(27.07,50.13){\circle*{2.00}} \put(39.07,50.13){\circle*{2.00}}
\put(51.07,50.13){\circle*{2.00}} \put(63.07,50.13){\circle*{2.00}}
\
\put(9.07,60.63){\circle*{2.00}} \put(21.07,60.63){\circle*{2.00}}
\put(33.07,60.63){\circle*{2.00}} \put(45.07,60.63){\circle*{2.00}}
\put(57.07,60.63){\circle*{2.00}}
\
\emline{3.07}{8.13}{1}{9.07}{18.63}{2}
\emline{15.07}{8.13}{3}{21.07}{18.63}{4}
\emline{27.07}{8.13}{5}{33.07}{18.63}{6}
\emline{39.07}{8.13}{7}{45.07}{18.63}{8}
\emline{51.07}{8.13}{9}{57.07}{18.63}{10}
\
\emline{15.07}{8.13}{11}{9.07}{18.63}{12}
\emline{27.07}{8.13}{13}{21.07}{18.63}{14}
\emline{39.07}{8.13}{15}{33.07}{18.63}{16}
\emline{51.07}{8.13}{17}{45.07}{18.63}{18}
\emline{63.07}{8.13}{19}{57.07}{18.63}{20}
\
\emline{9.07}{18.63}{21}{15.07}{29.13}{22}
\emline{21.07}{18.63}{23}{27.07}{29.13}{24}
\emline{33.07}{18.63}{25}{39.07}{29.13}{26}
\emline{45.07}{18.63}{27}{51.07}{29.13}{28}
\emline{57.07}{18.63}{29}{63.07}{29.13}{30}
\
\emline{9.07}{18.63}{31}{3.07}{29.13}{32}
\emline{21.07}{18.63}{33}{15.07}{29.13}{34}
\emline{33.07}{18.63}{35}{27.07}{29.13}{36}
\emline{45.07}{18.63}{37}{39.07}{29.13}{38}
\emline{57.07}{18.63}{39}{51.07}{29.13}{40}
\
\emline{3.07}{29.13}{41}{9.07}{39.63}{42}
\emline{15.07}{29.13}{43}{21.07}{39.63}{44}
\emline{27.07}{29.13}{45}{33.07}{39.63}{46}
\emline{39.07}{29.13}{47}{45.07}{39.63}{48}
\emline{51.07}{29.13}{49}{57.07}{39.63}{50}
\
\emline{15.07}{29.13}{51}{9.07}{39.63}{52}
\emline{27.07}{29.13}{53}{21.07}{39.63}{54}
\emline{39.07}{29.13}{55}{33.07}{39.63}{56}
\emline{51.07}{29.13}{57}{45.07}{39.63}{58}
\emline{63.07}{29.13}{59}{57.07}{39.63}{60}
\
\emline{9.07}{39.63}{61}{15.07}{50.13}{62}
\emline{21.07}{39.63}{63}{27.07}{50.13}{64}
\emline{33.07}{39.63}{65}{39.07}{50.13}{66}
\emline{45.07}{39.63}{67}{51.07}{50.13}{68}
\emline{57.07}{39.63}{69}{63.07}{50.13}{70}
\
\emline{9.07}{39.63}{71}{3.07}{50.13}{72}
\emline{21.07}{39.63}{73}{15.07}{50.13}{74}
\emline{33.07}{39.63}{75}{27.07}{50.13}{76}
\emline{45.07}{39.63}{77}{39.07}{50.13}{78}
\emline{57.07}{39.63}{79}{51.07}{50.13}{80}
\
\emline{3.07}{50.13}{81}{9.07}{60.63}{82}
\emline{15.07}{50.13}{83}{21.07}{60.63}{84}
\emline{27.07}{50.13}{85}{33.07}{60.63}{86}
\emline{39.07}{50.13}{87}{45.07}{60.63}{88}
\emline{51.07}{50.13}{89}{57.07}{60.63}{90}
\
\emline{15.07}{50.13}{91}{9.07}{60.63}{92}
\emline{27.07}{50.13}{93}{21.07}{60.63}{94}
\emline{39.07}{50.13}{95}{33.07}{60.63}{96}
\emline{51.07}{50.13}{97}{45.07}{60.63}{98}
\emline{63.07}{50.13}{99}{57.07}{60.63}{100}
\
\put(3.07,8.13){\line(1,0){12.00}}
\put(15.07,8.13){\line(1,0){12.00}}
\put(27.07,8.13){\line(1,0){12.00}}
\put(39.07,8.13){\line(1,0){12.00}}
\put(51.07,8.13){\line(1,0){12.00}}
\
\put(9.07,18.63){\line(1,0){12.00}}
\put(21.07,18.63){\line(1,0){12.00}}
\put(33.07,18.63){\line(1,0){12.00}}
\put(45.07,18.63){\line(1,0){12.00}}
\
\put(3.07,29.13){\line(1,0){12.00}}
\put(15.07,29.13){\line(1,0){12.00}}
\put(27.07,29.13){\line(1,0){12.00}}
\put(39.07,29.13){\line(1,0){12.00}}
\put(51.07,29.13){\line(1,0){12.00}}
\
\put(9.07,39.63){\line(1,0){12.00}}
\put(21.07,39.63){\line(1,0){12.00}}
\put(33.07,39.63){\line(1,0){12.00}}
\put(45.07,39.63){\line(1,0){12.00}}
\
\put(3.07,50.13){\line(1,0){12.00}}
\put(15.07,50.13){\line(1,0){12.00}}
\put(27.07,50.13){\line(1,0){12.00}}
\put(39.07,50.13){\line(1,0){12.00}}
\put(51.07,50.13){\line(1,0){12.00}}
\
\put(9.07,60.63){\line(1,0){12.00}}
\put(21.07,60.63){\line(1,0){12.00}}
\put(33.07,60.63){\line(1,0){12.00}}
\put(45.07,60.63){\line(1,0){12.00}}
\
\put(9.07,12.13){\circle{2.00}} \put(21.07,12.13){\circle{2.00}}
\put(33.07,12.13){\circle{2.00}} \put(45.07,12.13){\circle{2.00}}
\put(57.07,12.13){\circle{2.00}}
\
\put(3.07,22.63){\circle{2.00}} \put(15.07,22.63){\circle{2.00}}
\put(27.07,22.63){\circle{2.00}} \put(39.07,22.63){\circle{2.00}}
\put(51.07,22.63){\circle{2.00}} \put(63.07,22.63){\circle{2.00}}
\
\put(9.07,33.13){\circle{2.00}} \put(21.07,33.13){\circle{2.00}}
\put(33.07,33.13){\circle{2.00}} \put(45.07,33.13){\circle{2.00}}
\put(57.07,33.13){\circle{2.00}}
\
\put(3.07,43.63){\circle{2.00}} \put(15.07,43.63){\circle{2.00}}
\put(27.07,43.63){\circle{2.00}} \put(39.07,43.63){\circle{2.00}}
\put(51.07,43.63){\circle{2.00}} \put(63.07,43.63){\circle{2.00}}
\
\put(9.07,54.13){\circle{2.00}} \put(21.07,54.13){\circle{2.00}}
\put(33.07,54.13){\circle{2.00}} \put(45.07,54.13){\circle{2.00}}
\put(57.07,54.13){\circle{2.00}}
\
\put(78.07,50.13){\circle*{2.00}} \put(84.07,54.13){\circle{2.00}}
\emline{90.07}{50.13}{101}{96.07}{60.63}{102}
\emline{78.07}{50.13}{103}{84.07}{60.63}{104}
\put(84.07,60.63){\vector(1,2){0.01}}
\put(84.07,60.63){\line(1,0){12.00}}
\put(78.07,50.13){\vector(1,0){12.00}}
\put(82.07,60.63){\makebox(0,0)[rt]{$\ba_2$}}
\put(91.07,50.63){\makebox(0,0)[lt]{$\ba_1$}}
\
\put(68.07,8.13){\line(1,0){8.00}} \put(69.07,8.13){\circle*{2.00}}
\put(68.07,12.13){\line(1,0){8.00}} \put(72.07,12.13){\circle{2.00}}
\put(68.07,16.13){\line(1,0){8.00}} \put(69.07,16.13){\circle*{2.00}}
\put(68.07,18.80){\line(1,0){8.00}} \put(72.07,18.80){\circle*{2.00}}
\put(68.07,22.80){\line(1,0){8.00}} \put(69.07,22.80){\circle{2.00}}
\put(68.07,26.80){\line(1,0){8.00}} \put(72.07,26.80){\circle*{2.00}}
\emline{69.07}{8.13}{105}{72.07}{12.13}{106}
\emline{72.07}{12.13}{107}{69.07}{16.13}{108}
\emline{72.07}{18.80}{109}{69.07}{22.80}{110}
\emline{69.07}{22.80}{111}{72.07}{26.80}{112}
\
\put(83.07,8.13){\line(1,0){15.00}} \put(84.07,8.13){\circle*{2.00}}
\put(93.07,8.13){\circle*{2.00}} \put(83.07,12.13){\line(1,0){15.00}}
\put(87.07,12.13){\circle{2.00}} \put(96.07,12.13){\circle{2.00}}
\put(83.07,16.13){\line(1,0){15.00}}
\put(84.07,16.13){\circle*{2.00}} \put(93.07,16.13){\circle*{2.00}}
\put(83.07,18.80){\line(1,0){15.00}}
\put(87.07,18.80){\circle*{2.00}} \put(96.07,18.80){\circle*{2.00}}
\put(83.07,22.80){\line(1,0){15.00}} \put(90.07,22.80){\circle{2.00}}
\put(83.07,26.80){\line(1,0){15.00}}
\put(87.07,26.80){\circle*{2.00}} \put(96.07,26.80){\circle*{2.00}}
\put(83.07,29.46){\line(1,0){15.00}}
\put(90.07,29.46){\circle*{2.00}}
\put(83.07,33.46){\line(1,0){15.00}} \put(84.07,33.46){\circle{2.00}}
\put(93.07,33.46){\circle{2.00}} \put(83.07,37.46){\line(1,0){15.00}}
\put(90.07,37.46){\circle*{2.00}}
\emline{84.07}{8.13}{113}{87.07}{12.13}{114}
\emline{87.07}{12.13}{115}{84.07}{16.13}{116}
\emline{93.07}{8.13}{117}{96.07}{12.13}{118}
\emline{96.07}{12.13}{119}{93.07}{16.13}{120}
\emline{87.07}{18.80}{121}{90.07}{22.80}{122}
\emline{90.07}{22.80}{123}{87.07}{26.80}{124}
\emline{90.07}{29.46}{125}{93.07}{33.46}{126}
\emline{93.07}{33.46}{127}{90.07}{37.46}{128}
\
\put(77.07,12.13){\vector(0,1){4.00}}
\put(77.07,12.13){\vector(0,-1){4.00}}
\put(77.07,22.80){\vector(0,1){4.00}}
\put(77.07,22.80){\vector(0,-1){4.00}}
\put(82.07,12.13){\vector(0,1){4.00}}
\put(82.07,12.13){\vector(0,-1){4.00}}
\put(82.07,22.80){\vector(0,1){4.00}}
\put(82.07,22.80){\vector(0,-1){4.00}}
\put(82.07,33.46){\vector(0,1){4.00}}
\put(82.07,33.46){\vector(0,-1){4.00}}
\put(79.57,12.13){\makebox(0,0)[cc]{$\delta$}}
\put(79.57,17.46){\makebox(0,0)[cc]{$\Delta$}}
\
\put(33.07,3.13){\makebox(0,0)[lt]{(a)}}
\put(83.07,47.13){\makebox(0,0)[lt]{(b)}}
\put(79.57,3.13){\makebox(0,0)[lt]{(c)}}
\
\end{picture}
\caption[]{\label{LAYER} (a) Perpendicular projection of the
monolayer S-M-S ($\circ=$M, $\bullet=$S). (b) The elementary cell of
the monolayer containing two overlapping sulfur and one transition
metal atom. (c) Single layers of the 2Hb and 3R polytypes.}
\end{figure}

The monolayer M-S-M \cite{WILSON} is consisted of the pair of the
paralel sulfur planes at the distance $\delta\approx 6$\AA, bisected
by the transition metal plane (Fig.~\ref{LAYER}). All three planes
are of the same trigonal lattice, with the basis vectors $\vec{a}_1$
and $\vec{a}_2$ of equal length $a_0\approx 3$\AA. While the sulfur
atoms in the different planes are exactly one above another, the
metal atoms are between the centers of the sufur triangles. Thus
elementary cell of the layer contains two S atoms at $(0,0)$ (at the
heights $0$ and $\delta$) and the transition metal atom at
$(1/3,1/3)$ (at the height $\delta/2$). This layer has trigonal
symmetry of the diperiodic group Dg78=p\=6m2 (symorphic, with the
isogonal group $\mathbf{D}_{3h}$; horizontal mirror plane coincides
with M-plane) \cite{EW}.

The monolayered tube $(n_1,n_2)$ is formed when this lattice is
rolled up in such a way that the {\em chiral vector} ${\vec
c}=n_1\vec{a}_1+n_2\vec{a}_2$ becomes a horizontal circle on the
tube. Analogously to the carbon nanotubes, the {\em chiral angle}
$\theta=\arctan{\frac{\sqrt{3}n_2}{2n_1+n_2}}$ is the angle between
the lattice vector $\vec{a}_1$ and the chiral vector. Again, the
chiral angles from the interval $[0,\pi /6]$ suffice for the
description of all the tubes, since those with the angles from
$[\pi/6,\pi/3)$ are their optical isomers (the other angles results
in the same tubes, due to symmetry of the 2D lattice). Tubes $(n,0)$,
having zero chiral angle are called {\em zig-zag} ($\cZ$ for short),
the armchair ($\cA$) tubes are $(n,n)$ ones (with $\theta =\pi /6$),
while all others are referred as the {\em chiral} ones ($\cC$,
$n_1>n_2>0$, with $\theta\in (0,\pi /6)$).

The finite thickness of the monolayer results in different
distortions of the interior and exterior parts of the tube. It is
proposed \cite{REMSKAR} that the transition metal cylinder is purely
rolled up, while the interior and exterior sulfur ones are
additionally shrunken and stretched, respectively, such that their
radii are less (greater) then the metallic one for $\delta/2$.
Although diameter, as well as many other quantitative properties
(such as elastic energy or the dispersion relations), of the tube
depends on the described distortions, the symmetry group, and
consequently the symmetry based properties considered here are not
affected by this.

The diameter of the non-distorted cylinder is $|\vec{c}|/\pi$ (the
length of the chiral vector is $\sqrt{n_1^2+n_1n_2+n_2^2}a_0$). Thus,
the interior $D_{\rin}$ (sulfur) diameter is
\begin{equation}\label{ED}
 D_{\rin}=\sqrt{n_1^2+n_1n_2+n_2^2}a_0-\delta,
\end{equation}
and it will be mainly used in the following. The outer diameter is
obviously $D_{\rout}=D_{\rin}+2\delta$.

The symmetry groups of the MS$_2$ tubes are found by the procedure
established for the carbon nanotubes \cite{YITR}. It is based on the
fact that roto-translational tube symmetries emerge from
translational periodicity of the unfolded 2D layer. The only
modification is the odd order of the diperiodic group principle axis
(due to the mutual displacement of the sulfur and metallic
sub-lattices), resulting in absence of the $U$ axis. Here, only the
results, obtained by an easy modification according to this
difference, will be given. To this end, at first the number of the 2D
lattice cells in the elementary cell of the tube $(n_1,n_2)$ is found
to be
\begin{equation}\label{Eq}
 q=2\frac{n^2_1+n_1n_2+n^2_2}{n\cR}=\frac{2|\vec{c}|^2}{\cR a^2_0},
\end{equation}
where $n$ is the greatest common divisor of $n_1$ and $n_2$, while
$\cR=3$ if $n_1-n_2$ is multiple of $3n$, and $\cR=1$ otherwise. Note
that $3q$ is the number of atoms in the elementary cell of the tube.

The symmetry group of the chiral tube $(n_1,n_2)$ is the first family
line group
\begin{eqnarray}\label{ELc}
 \bL_\cC&=&\bT_q^r(a)\bC_n,\quad a=\sqrt{\frac{3q}{2n\cR}}a_0,\\
 &r&=\frac{n_1+2n_2-(\frac{n_2}{n})^{\varphi(\frac{n_1}{n})-1}q\cR}
  {n_1{\cal R}}\pmod{\frac{q}{n}}\nonumber
\end{eqnarray}
(the Euler function $\varphi(x)$ gives the number of coprimes not
greater than $x$). Each cylinder is one orbit (the set of atoms
generated from one of them by the group transformation) of the $a_1$
type \cite{LG}. The interior sulfur cylinder is used to fix the
coordinate system, choosing the orbit representative atom on the $x$
axis. Thus the cylindrical coordinates of the orbit representatives
are
\begin{equation}\label{Ecoo}
(\frac{D_{\rin}}{2},0,0),\quad(\frac{D_{\rin}+\delta}{2},\phi_0,z_0),\quad
(\frac{D_{\rin}}{2}+\delta,0,0),
\end{equation}
for interior sulfur, metal and outer sulfur, respectively. Here,
$\phi_0 =\pi\frac{n_1+n_2}{n_1^2+n_1n_2+n_2^2}$,
$z_0=\frac{a_0}{2\sqrt{3}}\frac{n_1-n_2}{\sqrt{n_1^2+n_1n_2+n_2^2}}$.
This orbit has trivial stabilizer (set of the group elements leaving
the representative atom invariant), meaning that at each cylinder the
atoms may be enumerated by the group elements. Precisely, the action
of the general element of the group (\ref{ELc}) on the point given by
the cylindrical coordinates is
\begin{equation}\label{Elgact}
 (C^r_q|\tfrac{q}{n}a)^tC^s_n\,
 (\rho,\phi,z)=(\rho,\phi+2\pi(\tfrac{rt}{q}+\tfrac{s}{n}),z+\tfrac{q}{n}ta).
\end{equation}
Thus, any atom in the cylinder (orbit) is given by its orbit
representative and pair $(t,s)$; its coordinates are given by
(\ref{Elgact}) with the corresponding representative coordinates from
(\ref{Ecoo}) taken for $(\rho,\phi,z)$.

The symmetry group of the zig-zag monolayered tube $(n,0)$ belongs to
the eighth line group family:
\begin{equation}\label{ELz}
  \bL_\cZ=\bL(2n)_nmc=\bT_{2n}^1(\sqrt{3}a_0)\bC_{nv}.
\end{equation}
Again, each cylinder is single $b_1$ orbit, with the coordinates of
orbit representatives given by (\ref{Ecoo}), where $\phi_0=\pi/n$ and
$z_0=\sqrt{3}a_0/6$. The stabilizer is $\bC_{1v}$, meaning that the
whole orbit can be generated by the chiral group (\ref{ELc}), with
$r=1$ and $q=2n$. The inner diameter is $D_{\rin}=na_0/\pi-\delta$.

For the armchair monolayered tube $(n,n)$ the symmetry is described
by the fourth line group family:
\begin{equation}\label{Ea}
\bL_\cA=\bL(2n)_n/m=\bT_{2n}^1(a_0)\bC_{nh}
\end{equation}
The cylinders are orbits of $b_1$ type generated by the atoms
positioned as in (\ref{Ecoo}), with $\phi_0=2\pi/3n$ and $z_0=0$. The
stabilizer is $\bC_{1h}$, and the orbit is effectively generated by
the chiral group again. The inner diameter is
$D_{\rin}=\sqrt{3}na_0/\pi-\delta$.

\subsection{Single-layered 2Hb and 3R polytype tubes}\label{Ssingle}
Both WS$_2$ and MoS$_2$ crystallize in two forms (space groups
$\mathrm{P}63/mmc$ and $\mathrm{R}3m$), differing in stacking of the
monolayers along the perpendicular direction. The single layer of the
hexagonal 2Hb polytype contains two monolayers positioned so that
sulfur (metal) atoms of the second monolayer are above the metal
(sulfur) atoms of the first one, i.e. the second monolayer is rotated
for $\pi/3$ (around perpendicular axis through S) and then translated
for $(\ba_1+\ba_2)/3$. Single layer of 3R polytype consists of three
monolayers: the second is translated for $(\ba_1+\ba_2)/3$ with
respect to the first (S atoms are moved above M atoms of the first
monolayer), while the third one is translated for $2(\ba_1+\ba_2)/3$
(its M atoms get above the S atoms of the first layer). In both cases
the distance between monolayers \cite{WILSON} is $\Delta\approx
2$\AA.

According to the experimental evidence, the walls of MS$_2$ tubes are
the 2Hb or 3R polytypes single layers, manifesting high
stretchability of the monolayers, \cite{REMSKAR}. The nanotubes
($D_{\rout}<0.1\,\mu{\rm m}$) grow in the hexagonal 2Hb polytype,
while the microtubes ($D_{\rout}>2\,\mu{\rm m}$) grow in rhombohedral
3R polytype, \cite{R99}. Despite different space groups of
crystalline polytypes, the layers symmetry is the same symorphic
diperiodic group Dg69=p3m1, with isogonal point group $\bC_{3v}$. In
comparison to the monolayer Dg78, it does not have the horizontal
mirror plane. When the layer is rolled up to the tube, such plain
produce no tube symmetry, meaning that the same line group is
obtained for the monolayered and both polytype single layer tubes of
the same chirality. Thus, the only difference is in the number of the
orbits out of which these systems are built: while the monolayered
tubes have $N=3$ orbits, the single-layered 2Hb tube consists of
$N=6$ orbits, and the single-layered 3R tube of $N=9$ orbits.

If the inner monolayer is taken as the referent one, i.e. its orbit
representative atoms have coordinates (\ref{Ecoo}), the atoms
generating the additional three orbits of the 2Hb single-layered tube
are positioned at (in the order from the inner to the outer orbits):
$$ (\frac{D^{(2)}_{\rin}}{2},\phi_0,z_0),\quad
(\frac{D^{(2)}_{\rin}+\delta}{2},0,0),\quad
(\frac{D^{(2)}_{\rin}}{2}+\delta,\phi_0,z_0),$$
where, $D^{(2)}_{\rin}=D_{\rin}+2\delta+2\Delta$ is the inner
diameter of the second monolayer. As for six additional orbit
representative atoms of 3R tubes one finds
$$ (\frac{D^{(2)}_{\rin}}{2},\phi_0,z_0),\quad
(\frac{D^{(2)}_{\rin}+\delta}{2},2\phi_0,2z_0),\quad
(\frac{D^{(2)}_{\rin}}{2}+\delta,\phi_0,z_0),$$
for the middle monolayer; the coordinates for the third monolayer
having the inner diameter $D^{(3)}_{\rin}=D_{\rin}+4\delta+4\Delta$,
are:
$$(\frac{D^{(3)}_{\rin}}{2},2\phi_0,2z_0),\quad
(\frac{D^{(3)}_{\rin}+\delta}{2},0,0),\quad
(\frac{D^{(3)}_{\rin}}{2}+\delta,2\phi_0,2z_0).$$
These data refer to the arbitrary chiral direction $(n_1,n_2)$. In
the special cases of zig-zag and armchair directions, the results are
obtained taking by substituting the values given for the
corresponding monolayers for $\phi_0$ and $z_0$.

\section{Symmetry determined properties of MS$_2$
nanotubes}\label{PROPERTIES}
Many physical properties are based on symmetry. Also, the symmetry
enables applications of the powerful group theoretical techniques in
solving various physical problems. In this paper, the presented
symmetry classification of the transition metal dichalcogenide tubes
is used to derive some of their physical properties determined by
symmetry, postponing the applications to various symmetry based
calculations (band structures, selection rules, etc.).
\subsection{Quantum numbers and band degeneracies}
Symmetry determines quantum numbers extensively used in almost all
problems (e.g. selection rules, bands assignation). Translational
periodicity is reflected in the conserved quasi-momentum $k$, taking
on the values from the 1D Brillouine zone $(-\pi/a,\pi/a]$. For the
armchair tubes, it can be taken from the irreducible domain
$[0,\pi/a]$ only \cite{ALTMANN-2}, since the horizontal mirror plane
$\sigma_h$ makes $k$ and $-k$ equivalent. The $z$-component of the
quasi-angular momentum $m$ is the quantum number due to the symmetry
of the isogonal rotations: since the order of the principle axis of
isogonal point group is $q$, $m$ runs over the integers from the
interval $(-\frac{q}{2},\frac{q}{2}]$ and characterizes the nanotube
energy bands. In the zig-zag tubes, the vertical mirror plane makes
$m$ and $-m$ equivalent, thus reducing this interval to its
nonnegative half. Since all of the nanotubes are non-symorphic, the
isogonal group is not a subgroup of the symmetry group; consequently
some of its rotations are not symmetries of the system at all, and
the quantum number $m$ is not a conserved quantity. Nevertheless,
these quantum numbers are frequently used, at the cost of more
complicated selection rules.

Alternatively, the conserved quantum numbers of helical
($\tilde{k}\in [0,q\pi/na]$) and pure angular quasi momenta
($\tilde{m}$, taking on the integer values from the interval
$(-\frac{n}{2},\frac{n}{2}]$) can be used. The former is related to
the momentum conjugated to the screw axis generated helix, and it
obviously incorporates a part of the angular momentum. The remaining
part $\tilde{m}$ is conserved, since it corresponds to the order $n$
principle axis of symmetry of the tubes.

The zig-zag (armchair) tubes have additional quantum number A/B (i.e.
+/-) of parity with respect to the vertical (horizontal) mirror
plane. Due to this extra symmetry the two dimensional irreducible
representations appear, i. e. the degeneracy of the energy bands can
be at most double or if the time-reversal symmetry of the
(spin-independent) Hamiltonian is taken into account, the maximal
degeneracy is fourfold, \cite{MLG}.

Concerning the chiral tubes, their energy bands are nondegenerate (as
far as the time-reversal symmetry is not included).

\subsection{Normal modes}
The classification of the vibrational bands depends only on the
symmetry group and the orbits out of which the system is formed. The
general classification has been performed already for all orbit types
of the line groups \cite{LG}. Since for the considered tubes are
decomposed to the orbits in the previous section, the results for
each of the found orbits are to be summed only.

All the tubes considered here consist of a finite number of the
orbits of the same type. The symmetry group and (henceforth) orbit
type depend on the chiral vector, not on the polytype (monolayer, 2Hb
or 3R): $a_1$ orbit for the chiral tubes and $b_1$ orbits for zig-zag
and armchair tubes (recall that these tubes have different
symmetries; i.e. while for the zig-zag tube the $b_1$ orbit
representative lay in the vertical mirror plane, for the armchair
tube it is in the horizontal plane). On the contrary, the number of
orbits is determined by the polytype: there is $N=3,6,9$ orbits in
the monolayered, 2Hb, 3R tubes, respectively. Thus the first step in
lattice dynamics study can be performed by the classification of the
vibrational bands, giving the band quantum numbers. Taking the
dynamical representation decomposition \cite{LG} for the found orbit
types, the phonons (normal modes) are classified. For the chiral
tubes one obtains
\begin{equation}\label{Edync}
 D^{{\rm dyn}}_{\cC}=3N\sum_{\tilde{k}}\sum_{\tilde{m}}{}_{\tilde{k}}A_{\tilde{m}},
\end{equation}
where the summation is performed over
$\tilde{k}\in(-\frac{q\pi}{na},\frac{q\pi}{na}]$ and integers
$\tilde{m}\in(-\frac{n}{2},\frac{n}{2}]$. It follows that there are
$3Nn$ vibrational bands assigned by the quantum number $\tilde{m}$.
The representations ${}_{\tilde{k}}A_{\tilde{m}}$ are one
dimensional, meaning that the bands are non-degenerate. Still, since
the Hamiltonian is real, the time reversal symmetry leads to the
double degeneracy of the bands over
$\tilde{k}\in(0,\frac{q\pi}{na}]$: the energies of the phonons with
quantum numbers $(\tilde{k},\tilde{m})$ and $(-\tilde{k},-\tilde{m})$
are equal.

The corresponding decompositions for the zig-zag and armchair tubes
are:
\begin{eqnarray}
 D^{{\rm dyn}}_{\cZ}&=&N\sum_{k\in(-\frac{\pi}{a},\frac{\pi}{a}]}
 \Bigl[2{}_kA_0+2{}_kA_n+{}_kB_0+\nonumber\\
 &+&{}_kB_n+N\sum_{m=1}^{n-1}{}_kE_m\Bigr],\label{EdynZA}\\
 D^{{\rm dyn}}_{\cA}&=&N\sum_{m=-n+1}^{n}({}_0A_m^-+2{}_0A_m^+)+
 \nonumber\\
 &+&3N\sum_{k\in(0,\frac{\pi}{a})}\sum_{m=-n+1}^n{}_kE_m+3N\sum_{m=1}^{n}{}_\pi E_m\label{EdynA},
\end{eqnarray}
Here, ${}_kA_m$ and ${}_kB_m$ correspond to non-degenerate $m$-bands
of different vertical mirror parity, while ${}_kE_m$ denotes the
double degenerate band. As for the armchair tubes, the second term
classifies the double degenerate bands, while the terms with $k=0$
and $k=\pi$ give insight in the bands topology (touching at the
Brillouine zone boundaries) and horizontal mirror parity of $k=0$
modes.

These decompositions can be used to simplify further vibrational band
calculations by standard methods. The obtained bands are
automatically assigned by good quantum numbers so that Raman and IR
spectra can be directly extracted by selection rules \cite{INY}, as
the relevant line group is found.
\subsection{Second rank tensors}

As the symmetry classification of the MS$_2$ tubes has been completed
the general matrix forms of the symmetric  polar and axial
second-rank tensors for systems with line group symmetry \cite{YITR}
can be used to predict many of the physical properties. Here the
optical activity of the isolated nanotubes will be analysed. In order
to avoid influence of the other effects, birefringence and anisotropy
of absorption, only the activity along the optical axis is discussed.
The optically active medium rotates the plane of polarization of a
linearly polarized incident light beam incoming along the optical
axis, defined by the unit vector $\mathbf{e}=(e_1,e_2,e_3)$, through
an angle $\sum_{i,j}e_ie_ja_{ij}$ per unit length, where $a_{ij}$
denote optical activity tensor elements. As this tensor is axially
symmetric second rank tensor, from the general axial tensor form
\cite{YITR} only the symmetric part is to be taken. The general form
of the dielectric tensor, i.e. a polar symmetric second-rank tensor
is needed to predict the spatial orientation of the optical axes.

The general form both of the tensor of the optical activity and of
the dielectric tensor for the chiral tubes with symmetry
$\bT^r_q\bC_n$, for $q>2$ what encounters all the realistic cases, is
given by the matrix:
 $$\tens{B}=\begin{pmatrix}b_{\bot}&0&0\\ 0&b_{\bot}&0\\ 0&0&b_{\Vert}\end{pmatrix}.$$
As the optical activity tensor does not vanish the chiral tubes are
expected to be optically active. The pair of equal eigenvalues of the
dielectric tensor singles out the tube axis as the optical axis. The
polarization plane is rotated through an angle proportional to the
value of $a_{\Vert}$. Since quasi-two dimensional transition metal
dichalchogenide layers are optically inactive, the value of
$a_{\Vert}$ continuously diminishes with the tube diameter.
As has been already mentioned, the tubes $(n_1,n_2)$ and $(n_2,n_1)$
are the optical isomers rotating the polarization of the incoming
light in the mutually opposite directions (through equal angles).

On the other side, zig-zag and armchair tubes are optically inactive
as due to the symmetry their second-rank axial tensors vanish.
\subsection{Potentials}

Potential $V(\br)$ in a nanotube must be invariant under the symmetry
transformations: $V(\br)=V(l^{-1}\br)$, for all $l\in\bL$, where
$\bL$ is the symmetry group of the tube considered. This property
leads to the quite restrictive conditions on the general form of the
potential function $V(\br)$. The translational (generated by $(I|a)$)
and rotational (generated by $C_n$) symmetries, manifested as
$V(\rho,\varphi,z)=V(\rho,\varphi,z-a)$ and
$V(\rho,\varphi,z)=V(\rho,\varphi-\varphi,z)$ enable the Fourier
expansion over coordinates $\varphi$ and $z$.

The screw axis generator $(C_q^r|\frac{na}{q})$ imposes the condition
$V(\rho ,\varphi ,z)=V(\rho ,\varphi -\frac{2\pi
r}{q},z-\frac{na}{q})$. This restricts the general form of the
potentials produced by a chiral tube to the Fourier expansion
\begin{equation}\label{FOURIER}
V_\cC(\rho ,\varphi ,z)=\sum_{\fracnul{K,M=-\infty}{(Mr=-K\,{\rm
mod}\frac{q}{n}})}^{+\infty}\,\alpha^M_K(\rho )e^{\ri nM\varphi}
e^{\ri (2\pi /a)Kz}
\end{equation}
(only the terms satisfying the braced equation are summed).

As for the zig-zag and armchair tubes $q=2n$ and $r=1$, this means
that the summation in (\ref{FOURIER}) is  performed only over the
terms with $K$ and $M$ of the same parity. The additional symmetries
are manifested as relations between the coefficients $\omega^M_K(\rho
)$, $\epsilon^M_K(\rho )$. The invariance of the potential of the
zig-zag (armchair) tubes under vertical (horizontal) mirror plane
$\sigma_v$ ($\sigma_h$) means $V(\rho ,\varphi ,z)=V(\rho ,-\varphi
,z)$ (i.e. $V(\rho ,\varphi ,z)=V(\rho ,\varphi ,-z)$); this leads to
the general form of the potential:
\begin{eqnarray}
 V_{\cZ}(\rho ,\varphi ,z)&=&\sum_{\fracnul{K=-\infty}{\mathrm{odd}}}^{\infty}
  \sum_{\fracnul{M=1}{\mathrm{odd}}}^{\infty}\,\omega^M_K(\rho )\cos{(Mn\varphi )}e^{\ri (2\pi
/a)Kz} +\nonumber\\
 &+&\sum_{\fracnul{K=-\infty}{\mathrm{even}}}^{\infty}
\sum_{\fracnul{M=0}{\mathrm{even}}}^{\infty}\, \epsilon^M_K(\rho
)\cos{(Mn\varphi)}e^{\ri (2\pi /a)Kz};\nonumber\\
V_{\cA}(\rho ,\varphi
,z)&=&\sum_{\fracnul{K=1}{\mathrm{odd}}}^{\infty}
\sum_{\fracnul{M=-\infty}{\mathrm{odd}}}^{\infty}\,\omega^M_K(\rho
)e^{\ri nM\varphi}\cos{(\frac{2\pi}{a}Kz)} +\nonumber\\
 &+&\sum_{\fracnul{K=0}{\mathrm{even}}}^{\infty}
  \sum_{\fracnul{M=-\infty}{\mathrm{even}}}^{\infty}\,\epsilon^M_K(\rho )e^{\ri
nM\varphi}\cos{(\frac{2\pi}{a}Kz)}.\nonumber
\end{eqnarray}
The obtained potentials can be further specified. When the Taylor
expansion of $\alpha^M_K(\rho )$, $\omega^M_K(\rho )$,
$\epsilon^M_K(\rho )$ is performed the terms of the same order in
$\rho$ should form the invariant polynomial of the relevant line
group, i.e. these coefficients are polynomials over the integrity
basis of the line group \cite{YITR}.

\section{Concluding remarks}
Geometry of the tubes is completely defined: line groups, number and
type of orbits, coordinates of the atom representing orbits are
specified generally, i.e.  for the tubes of arbitrary chirality. In
addition to the rotational, translational and screw axes symmetry of
the $(n_1,n_2)$, $0<n_2<n_1$ tubes, the armchair and zigzag tubes are
invariant under reflection in the mirror planes: the horizontal and
the vertical one, respectively. This leads to the parities of their
quantum states which now can be degenerate in contrast to the chiral
nanotube quantum states which (as far as the time-reversal symmetry
is not considered) are non-degenerate. Thus, the complete symmetry
groups of the chiral, armchair and zig-zag tubes are:
$$\bL_{\cC}=\bT_q^r\bC_n,$$
$$\bL_{\cA}=\bL_{\cC}+\sigma_h\,\bL_{\cC}=\bT_q^r\bC_{nh},$$
$$\bL_{\cZ}=\bL_{\cC}+\sigma_v\,\bL_{\cC}=\bT_q^r\bC_{nv}.$$
Note that the parameters $q$ and $r$ of the helical group have the
same form as the ones of the corresponding single-wall carbon
nanotubes \cite{YITR}.

The derived symmetry classification of the inorganic MS$_2$ nanotubes
is used to discuss some of their physical properties. First, the good
quantum numbers, characterizing the nanotube quantum states are
found: the translational, helical and rotational symmetries are
reflected in the conserved quasi-momenta $k$, $\tilde{k}$ and
$\tilde{m}$. The isogonal angular momentum $m$ is also used in
combination with $k$, although it is not a conserved quantity. The
ranges of these momenta are given. The vertical (horizontal) mirror
plane of zig-zag (armchair) tubes introduces the additional parity
quantum numbers denoted by $A$ and $B$ ($+$ and $-$) for even and odd
states, respectively.

The above enumerated quantum numbers are used in assignation and a
priori degeneracy prediction of electronic and vibrational bands. No
geometry caused degeneracy is expected for the chiral tubes, and
while for the zig-zag and armchair tubes double degeneracy in the
interior of the irreducible domain is obligate. Besides the time
reversal symmetry, which, when appropriate, doubles these
degeneracies, the diameter of the tube is also relevant in this
context. Namely, the diperiodic symmetry of the single 2Hb and 3R
layers (as well as that of monolayer) imposes six fold degeneracy of
the bands in the interior of 2D Brillouine zone. Thus, for the large
diameter tubes several (six for chiral and three for achiral ones)
bands are to become quasi degenerate. This effect should be
observable in microtubes.

Also, the tensors of optical activity and dielectric permeability are
studied. The optical activity of the chiral tubes and inactivity of
the arm-chair and zig-zag ones is demonstrated. In fact, these
results are typical, meaning that the same forms must be obtained for
any symmetric axial (polar) second rank tensor.

Finally, the general forms of the potentials in nanotubes are given,
thus enabling the generalization of the Bloch theorem to the line
group symmetry: by multiplication of the invariant potential
functions with irreducible representation matrix elements all the
symmetry allowed covariant and wave functions can be obtained.

\begin{acknowledgement} The authors would like to thank Dr. Maja Rem\v skar (J. Stefan
Institute, Ljubljana) and Professor Francis L\'evy (EPFL, Institut de
Physique Appliqu\'e) for helpful communications. One of us (IM) is
verrry thankful to Professor Werner Amrein (Universit\'e de Gen\`eve,
D\'epartement de Physique Th\'eorique) for providing her with the
opportunity to initiate this research.\end{acknowledgement}



\begin{thebibliography}{99}

\bibitem{WILSON}
J. A. Wilson and A. D. Yoffe, {\em Adv. Phys.} {\bf 18}  (1969) 193.

\bibitem{TENNE}
 R. Tenne, L. Margulis, M. Genut and G. Holdes, {\em Nature} {\bf 360}, (1992) 444;
 Y. Feldman, E. Wasserman, D. J. Srolovitz and R. Tenne, {\em Science} {\bf 267}, (1995) 222;
 R. Tenne, {\em Adv. Mater.} {\bf 7}, (1995) 965;
 G. L. Frey, S. Elani, M. Homyonfer, Y. Feldman, R. Tenne, {\em Phys. Rev.} {\bf B 57}, (1998) 6666.

\bibitem{R68}
 M. Rem\v skar, Z. \v Skraba, F. Cl\'eton, R. Sanjin\'es and F. L\'evy,
  {\em Surf. Rev. Lett.} {\bf 5}, (1998) 423;
  {\em Appl. Phys. Lett.} {\bf 69}, (1996) 351.

\bibitem{YITR}
 M. Damnjanovi\'c, I. Milo\v sevi\'c, T. Vukovi\'c and R. Sredanovi\'c,
  {\em Phys. Rev. B} {\bf 60}, (1999) 2728;
  {\em J. Phys. A} {\bf 32}, (1999) 4097.

\bibitem{LG}
 I. Milo\v sevi\' c and M. Damnjanovi\' c, {\em Phys. Rev.} {\bf B 47}, (1993) 7805;
 I. Milo\v sevi\' c, A. Damjanovi\' c and M. Damnjanovi\' c,
 in {\it Quantum Mechanical Simulation Methods for Studying Biological Systems},
   eds. D. Bicout and M. Field, Ch. XIV,
   (Springer-Verlag Berlin Heidelberg \& Les Editions de Physique Les Ulis, 1996).

\bibitem{ALTMANN-2}
 S. L. Altmann, {\em Band Theory of Solids, An Introduction from the
   Point of View of Symmetry} (Clarendon Press, Oxford, 1991).

\bibitem{EW}
 E. A. Wood, {\em 80 Diperiodic Groups in Three Dimensions}, Bell
  System Monograph No.4680 (1964);
 I. Milo\v sevi\' c I, M. Damnjanovi\' c, B. Nikoli\'c and M. Kr\v cmar,
{\em J. Phys. A} {\bf 31}, (1989) 3625.

\bibitem{REMSKAR}
 M. Rem\v skar, Z. \v Skraba, C. Ballif, R. Sanjin\'es, F. L\'evy,
  {\em Surf. Sci} {\bf 433-435}, (1999) 637.

\bibitem{R99}
 M. Rem\v skar, Z. \v Skraba, R. Sanjin\'es, F. L\'evy,
  {\em Appl. Phys. Lett.} {\bf 74}, (1999) 3633.

\bibitem{MLG}
 M. Damnjanovi\'c, I. Milo\v sevi\'c and M. Vuji\v ci\'c, {\em Phys. Rev.} B {\bf 39}, (1989) 4610;
 M. Damnjanovi\'c and I. Milo\v sevi\'c, {\em ibid.} {\bf 43}, (1991) 13482.

\bibitem{INY}
 I. Kirschner, C. Mészaros and R. Laiho, {\em Eur. Phys. J. B} {\bf 2}, (1998) 191;
 M. Damnjanovi\'c, I. Bo\v zovi\'c and N. Bo\v zovi\'c, {\em J.  Phys. A} {\bf 16}, (1983) 3937.

\end{thebibliography}
\end{document}